\documentstyle[epsfig]{elsart}

\begin{document}

\vspace*{1cm}

\begin{frontmatter}

\title{LOCALIZED DOMAINS OF DISORIENTED CHIRAL CONDENSATES}

{B.K. Nandi$^1$, T.K. Nayak$^2$, B. Mohanty$^1$, 
        D.P. Mahapatra$^1$ and Y.P. Viyogi$^2$ }

\address{$^1$Institute of Physics, Bhubaneswar 751005, India }
\address{$^2$Variable Energy Cyclotron Centre, 1/AF Bidhan Nagar,
         Calcutta 700064, India }

\date{\today}

\maketitle

\bigskip

\begin{abstract} 
    A new method to search for localized domains of
    disoriented chiral condensates (DCC) has been proposed
    by utilising the ($\eta-\phi$)
    phase space distributions of charged particles and photons. 
    Using the discrete wavelet transformation (DWT) analysis
    technique, it has been found that the presence of DCC
    domains broadens the distribution of wavelet coefficients
    in comparison to that of normal events. Strength contours 
    have been derived from the differences in rms deviations 
    of these distributions by taking into account the size of DCC 
    domains and the probability of DCC production in 
    ultra-relativistic heavy ion collisions. This technique can be 
    suitably adopted to experiments measuring multiplicities 
    of charged particles and photons. 
    
\end{abstract}


\end{frontmatter}



\bigskip

   Disoriented chiral condensates (DCC) have been predicted
   to be formed in  high energy hadronic and nuclear collisions
   when the chiral symmetry is temporarily restored at high temperatures.
   As the matter cools and expands, the vacuum may
   relax into a state that has an orientation different from the
   normal vacuum. This may lead to the formation of localized  domains of DCC,
   having an excess of low momentum
   pions in a single direction in isospin space 
   \cite{ans1,blai1,bjor1,raj1,asakawa}. 
   It has been estimated \cite{ans1} 
   that the neutral pion fraction, $f$, corresponding to the 
   DCC domain, follows the probability distribution:
\begin{equation}
 P(f) =  \frac{1}{2\sqrt{f}} {\rm ~~~~~~~~~~where~~~~} f = N_{\pi^0}/N_{\pi}
\end{equation}

   The formation of DCC domains gives rise to isospin
   fluctuation, where the neutral pion fraction can deviate
   significantly from 1/3, the value for uncorrelated emission of pions. 
   This would lead to 
   event-by-event fluctuation in the 
   number of charged particles and photons in a given phase space,
   since majority of the photons originate from $\pi^0$ decay and 
   the contents of  charged particles are mostly charged pions.

   The experimental observation of DCC depends on various factors, 
   such as the probability 
   of occurrence of DCC in a reaction, number of possible DCC domains in 
   an event, size of the domains, number of pions emitted from the domains 
   and the interaction of the DCC pions with rest of the system.
   However, not much theoretical guidance regarding  these factors is 
   available \cite{ans1,blai1,bjor1,raj1,asakawa,ishihara,horm1,randrup}. 
   While cosmic ray observations of Centauro and anti-Centauro
   events \cite{jacee} remain unexplained, the results of laboratory
   experiments  obtained so far \cite{wa98dcc,minimax}
   have not ruled out the presence of large fluctuations in the
   relative number of charged particles and photons in localized regions
   of ($\eta-\phi$) phase space. 
   The method of discrete wavelet transformation (DWT) \cite{huang}
   has been shown to be quite powerful to search for this 
   type of fluctuation \cite{nandi,qm97dcc,dccflow}.
   In this article, a new line of analysis has been employed 
   where the DWT technique has proved to be once again useful
   to reliably extract information regarding the size 
   and the frequency of occurrence of DCC domains. 

   The DWT technique has the beauty of 
   analyzing a distribution of particles at different length scales with 
   the ability of finally picking up the right scale at which there is a 
   fluctuation.
   This technique has been suitably adopted to search for bin to bin
   fluctuation in the charged particle and photon multiplicity distributions.
   The input to the DWT analysis is a sample distribution
   function at the highest resolution scale, $j_{max}$, 
   where the total number of bins is $2^{j_{max}}$.
   The sample function has been chosen to be:
\begin{equation}
    f^\prime=\frac{N_\gamma}{N_\gamma+N_{\rm ch}}
\end{equation}
   where $N_\gamma$ and $N_{\rm ch}$ are the multiplicities of photons and
   charged particles, respectively. The bins are made in $\phi$-space and
   analysis has been carried out using $D-$4 wavelet basis.
   The output consists of a set of wavelet or father function 
   coefficients (FFCs) at each scale, from $j=0$,...,($j_{max}-1)$. 
   The FFCs at a given scale carry information about the 
   degree of fluctuation at higher scales.
   Due to the completeness and orthogonality of the DWT basis, there is
   no information loss at any scale.



   The calculations for the present work have been performed for
   central (b $<$ 3 fm) 
   Pb+Pb collisions at the CERN-SPS energy of 158$\cdot$A GeV 
   using VENUS (version 4.12) event generator \cite{VENUS},
   which has been widely used to describe heavy ion collisions at
   the SPS energies. 
   All the particles with short lifetime are allowed to decay except $\pi_0$'s.
   In the absence of a reliable dynamical model where the effect of the
   formation of DCC could be simulated at an early stage of the reaction,
   the DCC pions are introduced at the freezeout stage. The
   goal is to observe the effect of DCC domains on the measured quantities,
   assuming DCC pions survive till the freezeout time. This assumption
   is supported by recent theoretical calculations \cite{koch}. 

   For simulations involving DCC domains, one needs to know the position
   of the domain, size of the domain in ($\eta,\phi$) and number of
   DCC pions within the domain. The DCC events are generated from
   VENUS events (referred to as 
``normal'' events)  by changing
   the relative population of charged and neutral pions \cite{nandi}
   in the domain according to the
   the following three  scenarios, varying from highly 
   optimistic to somewhat realistic cases:

\begin{itemize}
\item Assumption (1) Most optimistic case: \\
         All the pions within the chosen domain are assumed to be of
         DCC origin. 

\item Assumption (2) A little towards reality: \\
         Since pions from DCC domains are of low $p_T$, we assume that 
         out of all the pions present within the chosen domain, only
         pions with $p_T <$ 300 MeV/c may be of DCC origin. 
         (Such low $p_T$ pions constitute about half of the total 
         number of pions within the given domain.)

\item Assumption (3) Somewhat realistic: \\
         This corresponds to enhanced pion production
         in case of DCC formation. Here, a number of pions with $p_T <$
         300 MeV/c, generated according to 
         DCC probability given in equation (1) and having uniform $p_T$
         distribution, are added within the chosen
         domain on top of the existing pions in a normal event. 
\end{itemize}
   For assumption (3), the number of additional pions
   depends on the size of the DCC domain and the energy density of the
   domain. 
   DCC domain sizes of the order of 3$-$4~fm in radius have been
   predicted by different theoretical models \cite{gavin1,gavin2,kapusta}.
   Assuming  the energy density available for pion 
   production within the DCC domain to be about 50~MeV/fm$^3$,
   the number of DCC pions is estimated to be 
   between 40 and 100. Keeping this in mind,
   simulations have been performed taking 60 and 100 additional
   low $p_T$ pions.


   In the present work we confine ourselves to a single DCC
   domain, which is assumed to overlap with a hypothetical detector 
   system measuring charged particles and photons
   within the pseudo-rapidity range $3.0\le\eta\le 4.0$ and with full
   azimuthal coverage. 
   We assume a DCC domain of size $\Delta\eta$=1 
(within the above pseudo-rapidity  window)   
   and  vary the size in $\phi$ (denoted as $\Delta\phi$)
   to see the effect on observed quantities.
   It is advantageous to study small domains so as to avoid the attenuation
   caused by the sampling of uncorrelated pions \cite{randrup}.

   In order to be compatible with experimental observables, all simulations
   have been carried out by taking detector effects, like
   acceptance and efficiency, into account. 
   The efficiency of the charged particle detector is assumed to be 95\%
   \cite{spmd,alicetp}. 
   Because of the inherent difficulties in photon measurement,
   and guided by the previous works \cite{alicetp,neural,alicenote,wa98nim}, 
   the efficiency of photon detection is taken to be 70\%. It is also known
   that charged particles sometimes give signals in the photon detector and 
   such a contamination in an experimental situation can be 
   large. In view of this, a 25\% charged 
   particle contamination has been included in the photon signal.
   For assumption~3, DCC pions are added by taking into account the
   two track resolution of a typical detector.
   The average number of hits on the charged particle and photon 
   detectors for 
   normal events are  430 and 370 respectively.


   A multi-resolution DWT analysis on the sample function,
   as given by equation (2), is carried out starting
   with $j_{max}$=5.
   FFC distributions are obtained at scales, $j=1-4$, 
   for each of the above assumptions using 20000 events. The
   distributions are shown in Fig.~1 for scale $j=1$.
   The solid histograms in all the panels represent the FFC distributions
   for normal events with detector effects taken into account.
   The dashed histograms show the corresponding FFC distributions
   when a DCC domain of size, $\Delta\phi=90^o$ and $3\le\eta\le 4$ is
   introduced in {\it all} the events. This constitutes an 
   ensemble of DCC events having 100\% probability of occurrence of DCC.
   It is observed that the distributions for normal events are gaussian 
   at all scales, whereas there is a broadening in each of these 
   distributions in the presence of DCC. 

   The rms deviation, $\xi$, of the FFC distribution
   for the ensemble of  DCC events  is maximum for assumption~1, 
   as shown in Fig.~1(a), and reduces
   for the assumption (2) and (3) as seen in Figs. 1(b) and 1(c),
   respectively. 
   The reason for this is that for assumption~1 all the pions in the
   domain undergo DCC type of transformation, whereas for assumption~2,
   only   low $p_T$ pions, which are about half of the total number of pions,
   are allowed to be of DCC origin.
   In case of assumption~3, the number of extra 
   pions of DCC type, which are added
   on the top of the existing pions in a normal event, constitute only 
   20\% or less of the total pion population, thus further reducing 
   the broadening.

   The presence of a DCC domain modifies the phase space 
   distributions of charged particles and photons, resulting in an increase 
   in the rms deviations of the FFC distributions. To quantify this effect 
   properly, we introduce a strength parameter, $str$ as:
\begin{equation}
 str = \frac{\sqrt{(\xi_{\rm DCC}^2 - \xi_{\rm normal }^2)}}{\xi_{\rm 
                                                               normal}}
\end{equation}
   where $\xi_{\rm normal}$ is the rms deviation for an ensemble of
 normal events and
   $\xi_{\rm DCC}$ is the rms deviation for an ensemble of DCC events.

   For each of the DCC formation scenarios, 
   DCC events are generated for a given  domain
   size $\Delta\phi$. An ensemble of DCC events is then produced by
   mixing different fractions of DCC events with normal events,
   as appropriate for different
   probabilities of occurrence of DCC. A set of such ensembles of events are
   generated by varying the size $\Delta\phi$.
   Strength values are derived for all the different  event ensembles,
   the number of events  being 20000 in each case.  
   Contours of equal strength values are plotted corresponding to
   different domain sizes and varying probabilities of occurrence.
   Fig.~2 shows such contours for assumption~1,
   the top and bottom panels representing the contours 
   at scales $j=1$ and $j=2$ respectively. 
   The strength values are as marked for each contour in the figure. 

   The lowest contour corresponds to a strength value below which the 
   method is not sensitive either to the change of
   domain size or to the probability of occurrence of DCC, i.e., 
   no distinction can be made between normal events and DCC type of
   events.
   Using the rms deviation of the FFC distribution of 
   normal events and the statistical error 
   associated with it, we obtain the lowest strength value to be 0.1.
   The higher strength contours correspond to the cases where the FFC
   distributions are  wider compared to those of the normal events. 
   The figure indicates that the method is sensitive upto very low
   values of the probability of occurrence of DCC events, $\sim$0.2\%, 
   for domain sizes $\ge 90^o$ in case of 
   $j=1$ and 40$-$60$^o$ for $j=2$. 
   No significant change in $\xi$ values for $j=3$ and 4
   was noticed for DCC type of events compared to those of
   normal events.

   Fig.~3 shows a similar set of strength contours for assumption~2. 
   For a given domain size and probability of occurrence 
   of DCC events, the strength values in these cases are smaller compared to
   the corresponding values for assumption~1.  The lowest contour now appears
   for 1\% probability of occurrence of DCC.

   Fig.~4 shows the strength contours at scales $j=1-4$
   for 
   assumption~3, where additional pions from a DCC domain are superimposed
   on top of the normal events as discussed before.
   For this simulation, the number of DCC pions added are kept fixed
   irrespective of the size of the DCC domain. This may be justified
   if the coherence of the emitted pions is maintained till freeze-out
   even for smaller domains.
   The solid and dashed contours  correspond to 100 and 60
   additional DCC pions, respectively.
   In most cases the strength values for 100 
   additional pions are larger compared to those for 60 pions (for a given 
   domain size, equal strength appearing at a higher value of the
 probability of occurrence of DCC in the later case). 
   It is observed that the strength contours get broader
   in terms of DCC domain sizes, as one goes from scale $j=4$ to $j=1$.
   In our simulation $j_{max}$ is taken to be 5, 
   which corresponds to $\sim 11^o$ in $\phi$. We 
   find that the extent of the contours at a given
   scale of $j$, are from $11^o$ to $180^o/(2^{j-1})$ with minima at
   about $90^o/(2^{j-1})$.
   The fluctuation at the highest scale is thus carried over to the lower
   scales. The information obtained at all scales have to be 
   combined to conclusively characterize a DCC domain.

 
    This novel technique can be adopted to experiments measuring
    multiplicities of charged particles and photons in overlapping
    parts of phase space
   \cite{wa98,alice}. 
    Using the measured phase space distributions of particles, data
    analysis may proceed via DWT method to obtain FFC distributions
    at various scales. Similar set of FFC distributions can be obtained
    by simulating the experimental situation, with varying DCC domain sizes,
    and probability of occurrence and taking the detector effects 
    into account. 
    The strength values can be calculated from data with respect to the 
    simulated events at different scales. These values  shall correspond 
    to certain  equal strength contours obtained from the simulated events 
    with DCC, at different scales. If the  contour corresponding
    to the strength value of the data lies above the lowest contour of the
    simulated
    events, obtained by taking all systematic and statistical errors into 
    account, this might be indicative of the presence of measurable DCC
    effects in the data. 
    The effect will be best observed when the domain
    size is smaller compared to the detector coverage in actual experiments.  
    By correlating the information derived from the strength contours
    at all scales 
    one can draw inference about 
    the size and the number of pions emitted from a localized 
    DCC domain present in the data and also about 
    the probability of occurrence of DCC.



    In summary, the difference in the rms deviations of the FFC distribution
    of DCC type of events with respect to normal events has been 
    exploited for characterizing the localized domains of DCC.
    The sensitivity of the DWT technique to pick up fluctuations has been
    well demonstrated. It has been found that by correlating the rms 
    deviations of the FFC distributions at different scales one can predict 
    the size and the location of such fluctuations. 
    The method uses three assumptions motivated from various 
    theoretical predictions, and has been able to show that the
    sensitivity of this method is best in the most realistic assumption~3 
    scenario. Because of very high particle multiplicities at RHIC and LHC,
one can go up in scale for DWT analysis, without encountering large 
bin-to-bin statistical fluctuations. This will enhance the sensitivity of the
method and will allow one to probe smaller domains.

\newpage

\vspace*{1cm}
\noindent {\Large \bf Figure Captions }

\rm

\noindent {\bf Figure 1.} 
The FFC distributions at scale $j=1$ for
VENUS events (solid histograms) and
DCC type of events (dashed histograms),
with DCC domain size $\Delta\phi=90^o$ and $3\le\eta\le 4$,
corresponding to
(a) Assumption~1, (b) Assumption~2, (c) Assumption~3.

\noindent {\bf Figure 2.} 
Strength contours at two different scales, $j=1$ and $j=2$
for DCC assumption~1. The abscissae shows the size of DCC domains in
$\Delta\phi$  and the ordinate is the probability
of occurrence of DCC domains.

\noindent {\bf Figure 3.}
Strength contours at two different scales, $j=1$ and $j=2$
for DCC assumption~2, where only low $p_T$ pions ($p_T<300$ MeV)
are assumed to be of DCC origin.

\noindent {\bf Figure 4.}
Strength contours at four resolution scales, $j=1$,2,3 and 4,
for DCC assumption~3. The solid curves show the contours for
100 low $p_T$ ($p_T<300$ MeV) additional DCC pions 
and the dashed curves are for 60 additional DCC pions.

\begin{figure}
\begin{center}
\epsfig{figure=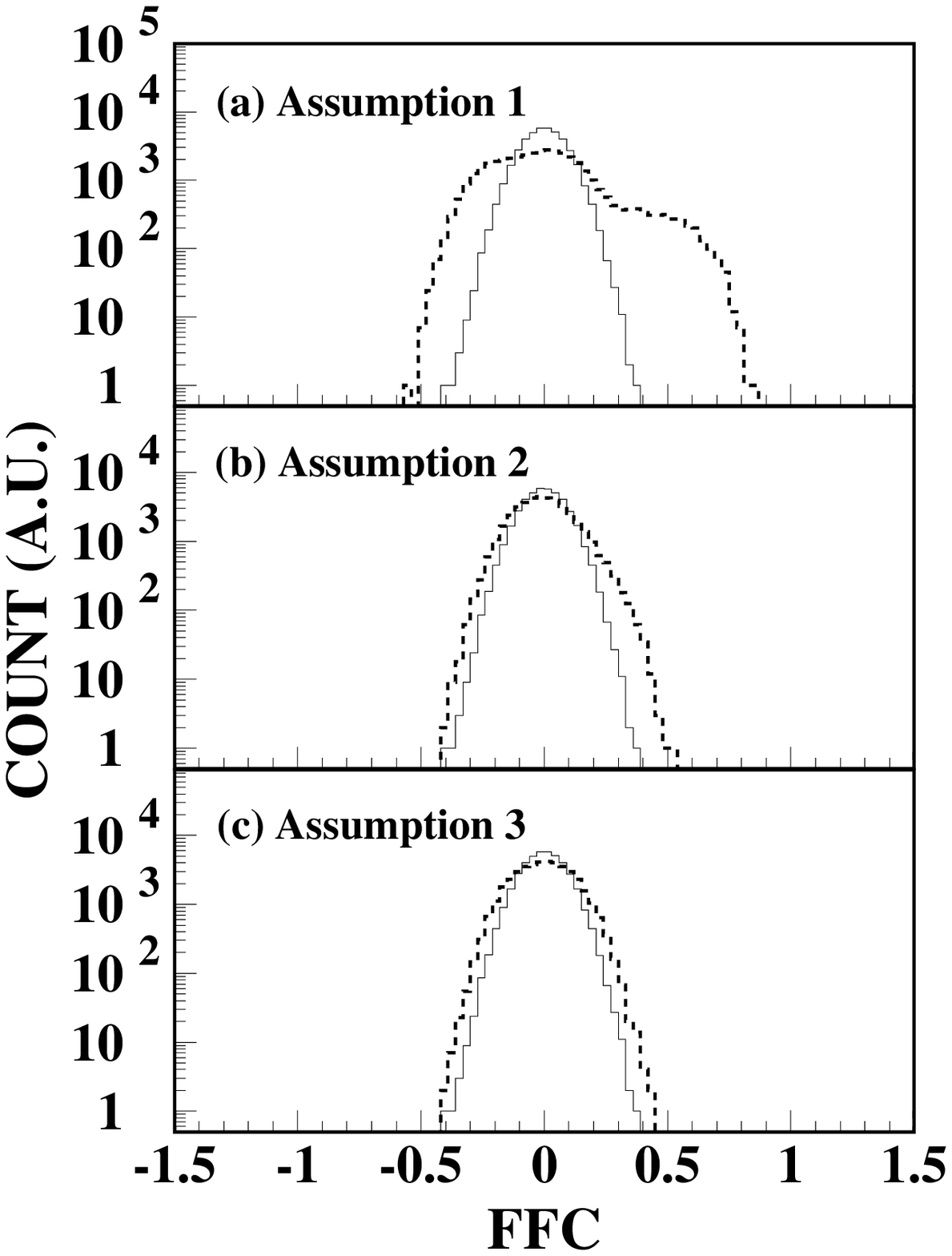,width=14cm}
\caption {\label{ffc1}
}
\end{center}
\end{figure}

\begin{figure}
\begin{center}
\epsfig{figure=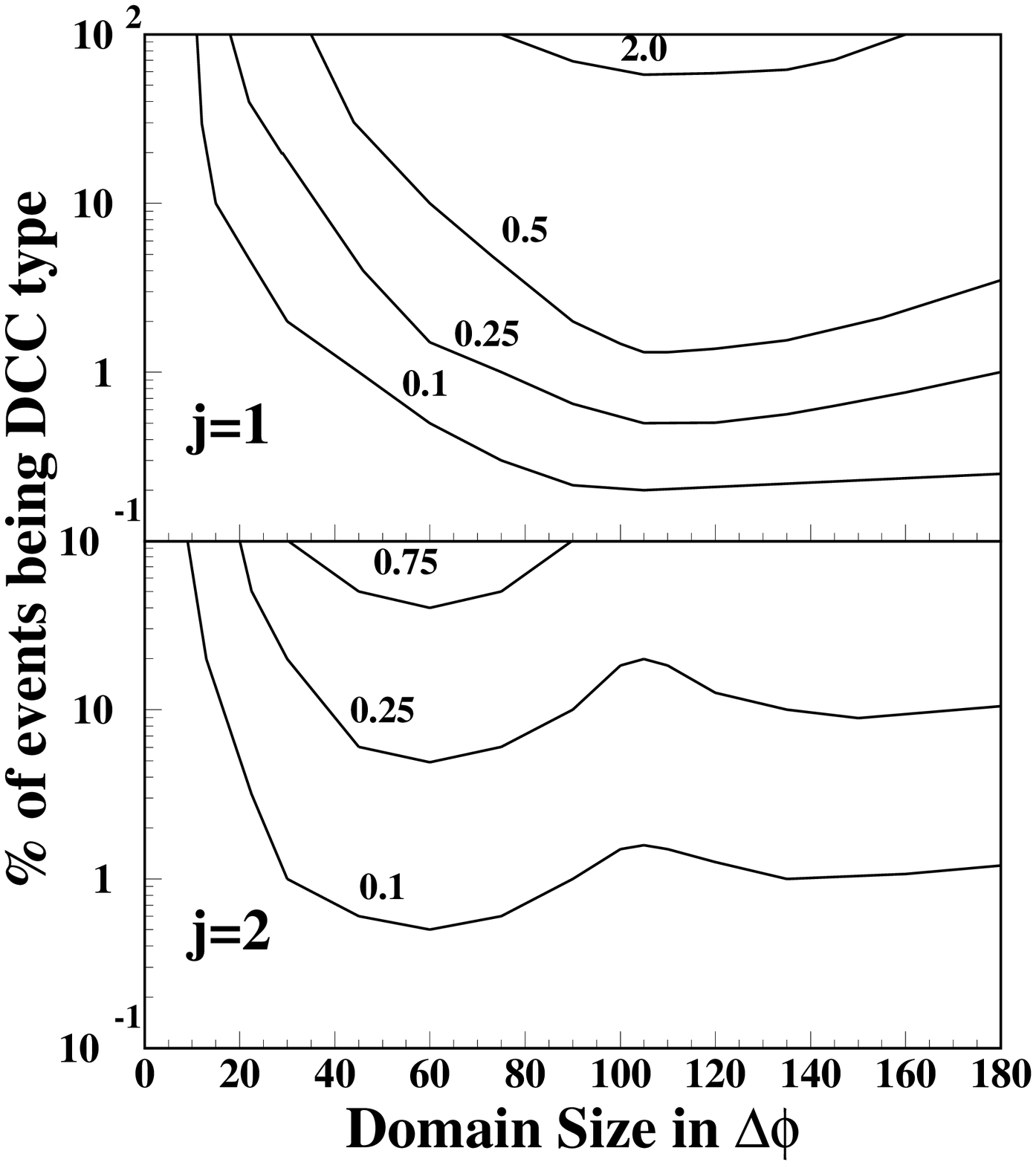,width=14cm}
\caption {\label{cont1}
}
\end{center}
\end{figure}

\begin{figure}
\begin{center}
\epsfig{figure=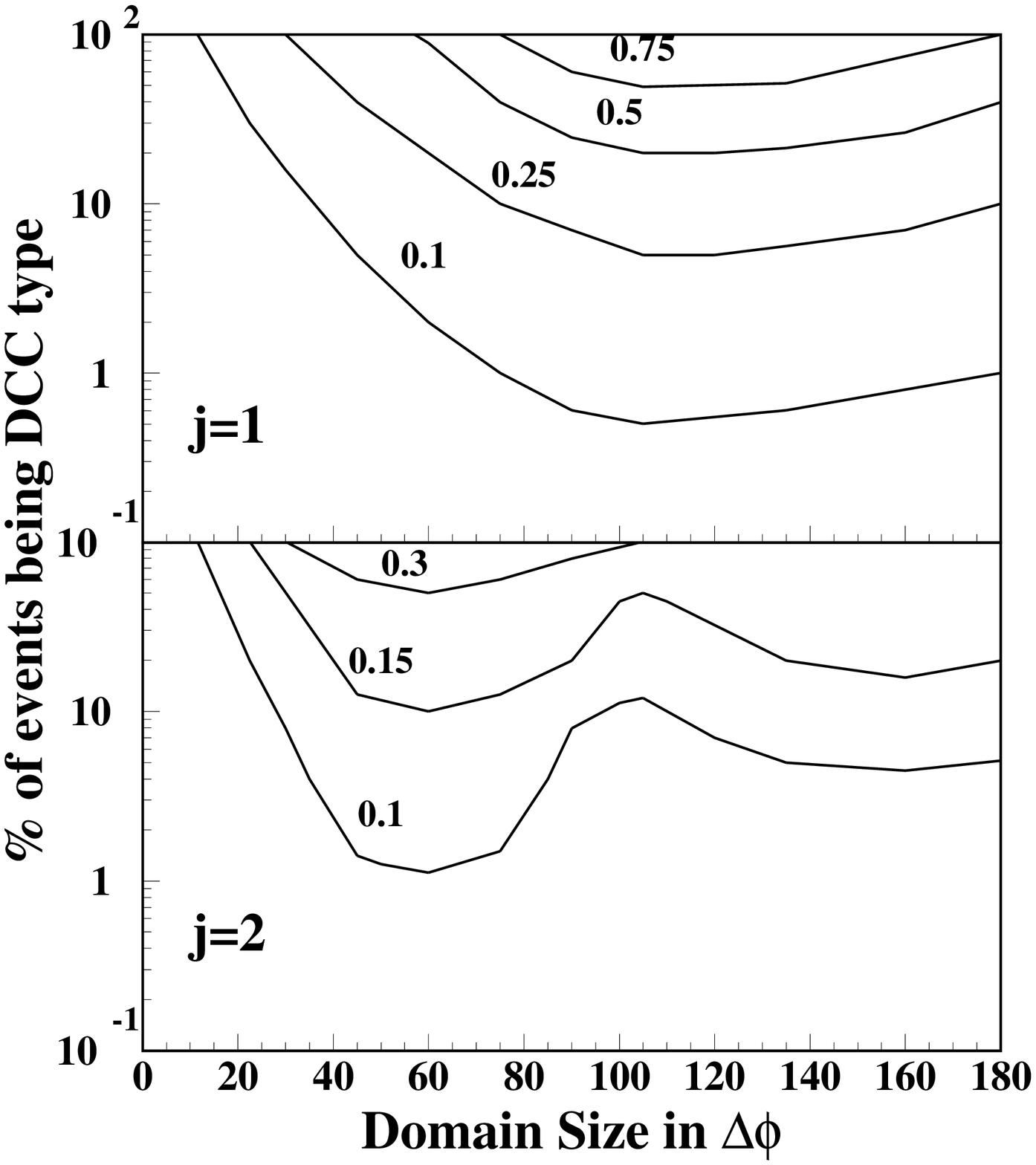,width=14cm}
\caption {\label{cont2}
}
\end{center}
\end{figure}

\begin{figure}
\begin{center}
\epsfig{figure=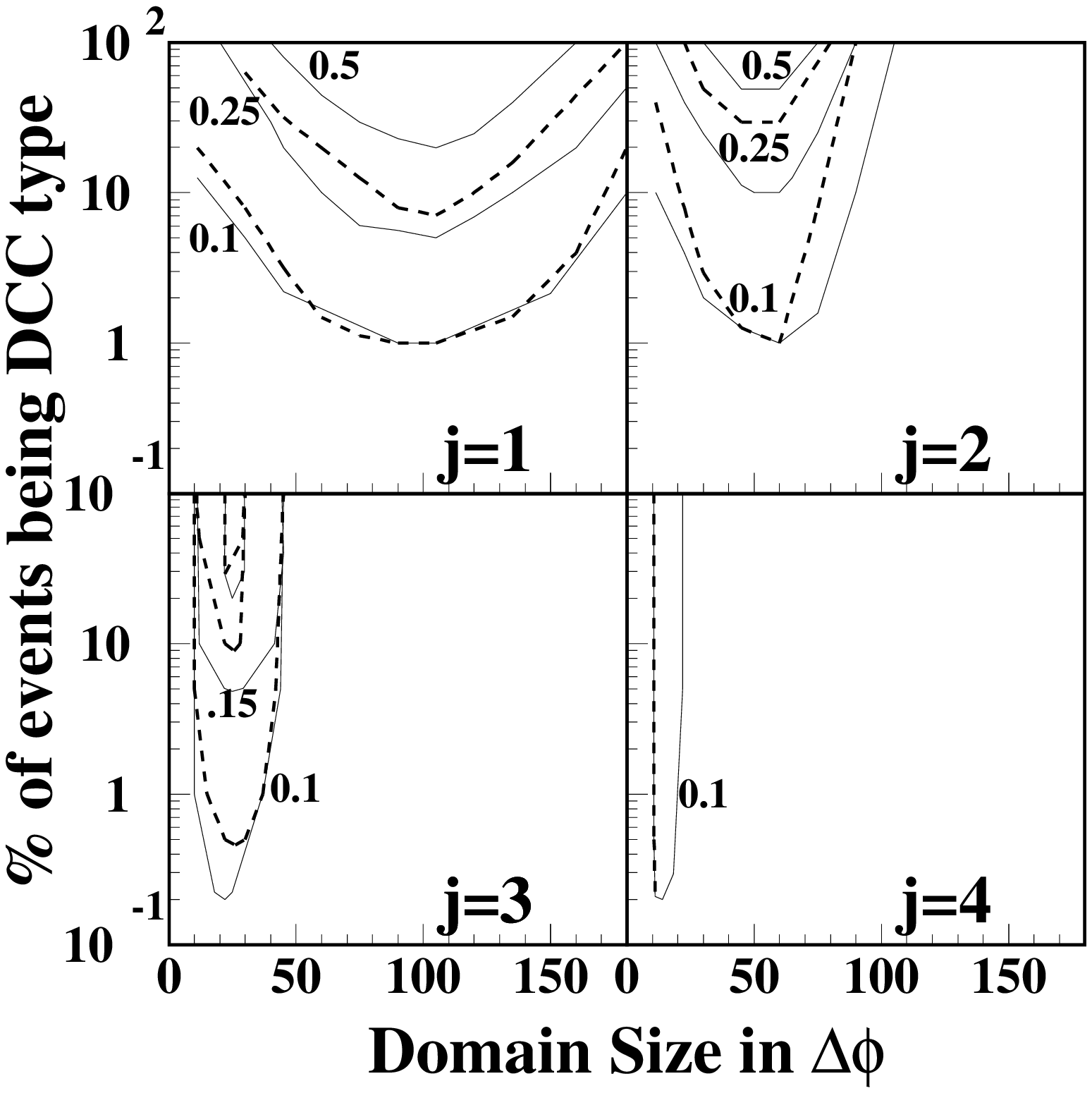,width=16.cm}
\caption {\label{cont3}
}
\end{center}
\end{figure}


\begin{thebibliography}{99}

\bibitem{ans1}      A.A. Anselm, M.G. Ryskin, 
                    Phys. Lett. {\bf   B266} (1991) 482.
\bibitem{blai1}     J. -P. Blaizot and A. Krzywicki, 
                    Phys. Rev. {\bf D46} (1992) 246.
\bibitem{bjor1}     J.D. Bjorken, Int. J. Mod. Phys. {\bf A7} (1992) 4189;
                    J.D. Bjorken, K.L. Kowalski, C.C. Taylor, ``Baked Alaska'',
                    SLAC-PUB-6109, Apr. 1993.
\bibitem{raj1}      K. Rajagopal, F. Wilczek, 
                    Nucl. Phys. {\bf B399} (1993) 395.
\bibitem{asakawa}  Masayuki Asakawa, Zheng Huang and Xin-Nian Wang,
                Phys. Rev. Lett. {\bf 74} (1995) 3126.


\bibitem{ishihara} M. Ishihara, M. Maruyama and F. Takagi,
                Phys. Rev. {\bf C57} (1998) 1440.
\bibitem{horm1} James Hormuzdiar and Stephen D.H. Hsu,
                Phys. Rev. {\bf C58} (1998) 1165.

\bibitem{randrup} Jorgen Randrup and Robert L. Thews,
                Phys. Rev. {\bf D56} (1997) 4392. 

\bibitem{jacee}   Proceedings of VIII International Symposium on Very High
                    Energy Cosmic Ray Interactions (Tokyo, Japan), 
                    24-30 July 1994; \\
                    C.M.G. Lattes, Y. Fujimoto, and S. Hasegawa, 
                    Phys. Rep. {\bf 65} (1980), 151; \\
                    J. Lord and J. Iwai, International conference on HEP 
                    (Dallas, 1992).

\bibitem{wa98dcc} M.M. Aggarwal et al., (WA98 Collaboration), Phys. Lett.
                  {\bf B420} (1998) 169.
\bibitem{minimax} T.C. Brooks et al., (Minimax Collaboration), 
                  Phys. Rev. {\bf D55} (1997) 5667.
\bibitem{huang}   Zheng Huang, Ina Sarcevic, Robert Thews and Xin-Nian Wang, 
                  Phys. Rev. {\bf D54} (1996) 750.
\bibitem{nandi}     B.K. Nandi (WA98 Collaboration), Proc. $3^{\it rd}$
                    Int'l. Conf. on Physics and Astrophysics of  
                    Quark-Gluon
                    Plasma (ICPA-QGP '97), ed. B. Sinha, D.K. Srivastava,
                    Y.P. Viyogi, Narosa Publ. House, New Delhi (1998)
                    p. 532.
\bibitem{qm97dcc}  T.K. Nayak et al., (WA98 Collaboration),
                   Quark Matter '97 proceedings, 
                   Nucl. Phys. {\bf A638} (1998) 249c.
\bibitem{dccflow}  B.K.~Nandi, G.C.~Mishra, B.~Mohanty, D.P.~Mahapatra
                   and T.K.~Nayak, Phys. Lett. {\bf B449} (1999) 109.
\bibitem{VENUS}    K. Werner, Phys. Rept. {\bf 232} (1993) 87.
\bibitem{koch}     James V. Steele and Volker Koch,
                   Phys. Rev. Lett. {\bf 81} (1998) 4096.
\bibitem{gavin1}    Sean Gavin, Andreas Gocksch and Robert D. Pisarski,
                Phys. Rev. Lett. {\bf 72} (1994) 2143.
\bibitem{gavin2}   S. Gavin and B. Muller, Phys. Lett. {\bf B329} (1994) 486.
\bibitem{kapusta} J.I. Kapusta and A.P. Vischer, Z. Phys. {\bf C75} 
                  (1997 ) 507.
\bibitem{spmd}    W.T.~Lin et al., 
                  Nucl. Instr. Methods {\bf A389} (1997) 415.
\bibitem{alicetp}   M.M. Aggarwal et al., {\it Internal Report},
                    VECC/EQG/99-01.
\bibitem{neural} S. Chattopadhyay, Z. Ahammed and Y.P. Viyogi,
                 Nucl. Instr. Methods {\bf A421} (1999) 558.
\bibitem{alicenote} Y.P. Viyogi, ALICE Internal note 98-52.
\bibitem{wa98nim} M.M. Aggarwal et al., {\it A Preshower Photon
                 Multiplicity Detector for the WA98 Experiment}, 
                 hep-ex/9807026, Nucl. Instr. and Methods, in press.

\bibitem{wa98}
 H.H. Gutbrod et al.,
 Proposal for a Large Acceptance Hadron and Photon
Spectrometer,  CERN-SPSLC-91-17, CERN-P-260 (1991).

\bibitem{alice} ALICE Technical Proposal, CERN/LHCC 95-71, LHCC/P3 (1995).


\end{thebibliography}
\end{document}